\documentstyle[12pt,aasms4]{article}
\slugcomment{Accepted by the Astronomical Journal}

\def\spose#1{\hbox to 0pt{#1\hss}}
\def\simlt{\mathrel{\spose{\lower 3pt\hbox{$\mathchar"218$}}
     \raise 2.0pt\hbox{$\mathchar"13C$}}}
\def\simgt{\mathrel{\spose{\lower 3pt\hbox{$\mathchar"218$}}
     \raise 2.0pt\hbox{$\mathchar"13E$}}}

\def\etal{{\rm et~al.~}}

\received{August, 2004}
\accepted{November, 2004}
\lefthead{Mould}
\righthead{Red thick disks}

\begin{document}
\title{Red thick disks of nearby galaxies}

\author{Jeremy Mould}
\affil{NOAO, PO Box 26732, Tucson AZ 85726-6732}
\authoraddr{E-mail: jrm@noao.edu}


\begin{abstract}
Edge on systems reveal the properties of disk galaxies as a function of
height, z, above the plane. Four local edge-on galaxies, that are close enough
to have been resolved into stars by the Hubble Space Telescope, show thick disks,
composed of a red stellar population, which is old and relatively metal rich.
Color gradients, $\Delta$(V-I)/$\Delta$z, are zero or slightly positive.
Favored models may have an explicit thick disk formation phase.
\end{abstract}


\section{Introduction}

Thick disks were discovered twenty five years ago (Gilmore and Reid 1983, Burstein 1979), but only 
the most basic constraints have subsequently been set on their origin in the structural
formation of galaxies. 
An understanding of thick disks lies at the heart of the set of physical
questions about the origin and evolution of disks in galaxies. The
shredding of satellite galaxies, the triggering of star formation in
disk gas by mergers, and the early evolution of disks in protogalaxies,
while they are still out of dynamical equilibrium,
are three key processes which must contribute to the thick disk fossil
record. Observations and simulations provide significant opportunities
to diagnose the relative importance of these three processes, and to
discover other processes.

Recent developments in theory and observation include the work of Abadi \etal (2003),
who point to a primary role for residual stars from merged dwarfs,
and Dalcanton, Yoachim, and Bernstein (2004), who find a critical mass
(or rotation velocity) below which star formation efficiency may be
markedly lower. Other, only slightly less recent work, that bears directly
on these disk formation and evolution processes, is by Dominguez-Tenreiro,
Tissera, and Saiz (1998) and Johnston, Sackett, and Bullock (2001).
Most recently, Brook \etal (2004) have explored early results from chemo-dynamical
cosmological simulations and identify thick disk formation with a z $>$ 2 epoch
of chaotic merging of multiple, peer disks of mixed gas and stars.

New insight into the disk component of galaxies finds that thick disks are prevalent, old, and red. Dalcanton \& Bernstein (2002) find these properties in surface photometry of an edge-on sample, and argue that they are the markers of an early epoch of merging experienced by all disk galaxies. If this is the case, the resolved stellar population of an edge-on nearby sample of disk galaxies should show both the characteristic age and metallicity profile of a post-halo epoch of star formation 6-12 Gyrs ago. 

A suitable sample for study of (I,V--I) color magnitude diagrams consists of the closest galaxies with axial ratios less than 0.25. Galaxies in NED,
whose redshift implies a distance less than 25\% that of the Virgo cluster  
according to Mould \etal (2000), and with A$_B~<$ 1, are ESO318-G3, 383-G91,
NGC 784, 1560, 4144, 4244, IC 3308, UGC 1281, 7321, 11583, and UGCA 442.  
Those with available WFPC2 data are NGC 784, 1560, UGC 1281, 7321, UGCA 442, and IC 3308. UGC 7321 (Matthews, Gallagher \& van Driel 1999) and IC 3308 are
not well resolved by WFPC2. In the present paper we analyze NGC 784, 1560,
UGC 1281, and UGCA 442. They span a range of rotation velocities from 95 to 
150 km/sec (21 cm line full width at 20\% peak). 

WFPC2 on HST reaches below the tip of the red giant branch (RGB) in these galaxies, allowing us to diagnose the star formation history of their thick disk stellar populations. The color of the RGB in the magnitude fainter than the tip will tell us the metallicity of the population (Da Costa \& Armandroff 1990).
Another diagnostic tool is the extended giant branches of Magellanic Cloud star clusters in this age range, in which the luminosity difference between the asymptotic giant branch (AGB) tip (M$_{bol,f}$) and the RGB tip (M$_{bol,TRGB}$) 
is a measure of age (Mould \& Aaronson 1982, Frogel, Blanco, \& Mould 1989), and the fuel burning theorem (Renzini \& Buzzoni 1987) tells us the star formation history over the epoch. 

This project is intended to lay foundations for a larger study of the formation epoch of local galaxies, which will complement the analysis of that epoch obtained by `looking back' at very distant galaxies. Freeman and Bland-Hawthorn (2002) offer the prospect that the chemical and kinematic evolution of the early history of disks may be deterministically unraveled in detailed spectroscopic studies with large telescopes. The present study is a start on the reconnaissance of the region from 0 to 5 Mpc to be explored in this way, using simple photometry of resolved stars.

\section{Color Magnitude Diagrams of four edge-on disk galaxies}

The Hubble Space Telescope archive contains WFPC2 images of four of the sample
galaxies listed in Table 1. The relevant observational parameters are
noted in columns (3) and (4).
Column (5) of Table 1 is the Galactic extinction in I magnitudes drawn from NED.
This database is also the source of the axial ratios (column 2), which quantify
the edge on morphology of the galaxies. Images of the galaxies made with standard STSDAS-IRAF cosmic ray rejection are reproduced in Figure 1 (not part
of the astro-ph version). 

Photometry of all four chips in both filters in all four galaxies was carried
out using DAOPHOT (Stetson 1987) and ALLSTAR (Stetson 1994). Stars were detected at the 2$\sigma$ level on the images, and WFPC2 point spread functions were taken from Stetson \etal (1998), following the practice of the Hubble Constant
Key Project (Kennicutt, Freedman, \& Mould 1995). Linking the frames was achieved trivially with DAOMASTER, since all images were undithered and single epoch.
The calibration and CTE corrections of Dolphin (2002) were adopted, as updated
http://purcell.as.arizona.edu/$\sim$andy/wfpc2\_calib
Aperture corrections were measured using bright stars cleaned with SUBSTAR.
The four color magnitude diagrams (CMDs) are shown in Figures 2--5.

In all four CMDs the upper AGB is prominent, but confined to the plane of the galaxy. In the giant branch luminosity function a rapid rise at I magnitudes
between 23 and 24 corresponds to the tip of the RGB. The `edge detector'
of Sakai \etal (1996) is then used to effectively differentiate the
luminosity function. 
The tip of the RGB is thereby detected as a peak (bottom left plots
in Figures 2b--5b), allowing the distance of each galaxy to be 
measured (Madore \& Freedman 1998 and references therein).
Bellazzini, Ferraro, \& Pancini (2001) find $$M_I^{TRGB} = 0.14[Fe/H]^2
+0.48[Fe/H] -3.66$$. Adopting --1 $<$ [Fe/H] $<$ --0.3, we calculate a
calibration uncertainty $\delta M_I^{TRGB}$ = $\pm$0.10 mag. Observational
uncertainty is explored in the Appendix by means of artificial star experiments.
Apparent distance moduli are reported in column (6), Table 1. The quoted
uncertainty reflects chemical, calibration, and observational uncertainties in M$_{bol,TRGB}$. 

The question arises whether the RGB tip distance may be affected
by internal reddening in the plane of each galaxy. 
Comparing the lower two panels
in Figure 3b, we see no difference in the the RGB tip in and off the plane of
NGC 1560 and 0.2 mag difference in UGCA 442 and UGC 1281.
The luminosity function statistics of NGC 784 makes
such a comparison noisy and difficult, but an upper limit of 0.6 mag
can be set. For galaxies of this morphological type (T $\simgt$ 6)
Han (1992) finds $\Delta$I = 0.33 $\pm$ 0.14 to 0.49 $\pm$ 0.22 mag for 
galaxies with this range of axial ratios. These values are probably an upper
limit on the extinction of the old red giants and only relevant to the plane
of the galaxies. As a possible systematic error, the in-and-off the plane 
difference in M$_{TRGB}$ is added directly to the distance error estimate in Table 1,
rather than in quadrature, as would be appropriate for random errors.

Other measurements of the distances of these galaxies are sparse. NGC 1560 is
a member of the Maffei group of galaxies according to Kraan-Korteweg (1985),
hereafter KK, who places NGC 1560 at 0.20 the distance of the Virgo cluster.
That yields 3.0 Mpc from the Virgo distance of Mould, Kennicutt, \& Freedman
(2000) which can be compared
with 3.3 $\pm$ 0.2 Mpc for our distance for NGC 1560. KK has UGCA 442 as a
member of the Sculptor group, 3.4 $\pm$ 0.3 Mpc away according to Aaronson and Mould (1983). Our UGCA 442 distance is 3.8 +0.8/-0.3 Mpc. The other
galaxies are field galaxies. UGC 1281 is approximately 0.3 the distance of Virgo
according to KK, or 4.5 Mpc, compared to our measured 4.2
+0.9/-0.3 Mpc. For NGC 784 the comparison is 5.4 (from KK) $cf.$ 3.0 +1.5/-0.2 Mpc.
With the exception of the last, the agreement is satisfactory for present purposes, but, as a caveat, we note that our distances are uncorrected for internal
extinction, and that alternative distances from Virgocentric flow models
are subject to random velocities that are a significant fraction of the
recession velocity.

\section{Vertical color gradients}

As a measure of a mean color for the giant branch in these CMDs, it is 
straightforward to define an error weighted average of the V--I colors
of stars within a broad color window of the RGB/AGB. The window was defined
by 0.80 $<~V-I~<$ 2.34 and 19 $<~I~<$ 26 mag. The lower limit on color excludes
main sequence stars and blue supergiants. The upper limit on magnitude
excludes red horizontal branch stars, but includes red supergiants.
This window is consistent for all 4 galaxies and for the models
described in the next section.
Figure 6 shows the color gradient in this $<V-I>$
normal to the projected plane of the disk. Since reddening might affect the
colors in the thin disk within a few hundred parsecs of the midplane, colors
are not plotted close to the plane. No correction has been made for
interstellar reddening in Figure 6. No correction is required for foreground
stars, which are few in number. For example, for NGC 1560, the lowest
latitude member of the sample, the Bahcall \& Soneira (1986)
model of the galaxy predicts 7 foreground stars in the color window defined 
above with 19 $<~V~<$ 27. The mean color is V--I = 1.6, based on a standard
(B--V,V--I) diagram, very similar to the color of the giant branch stars 
themselves.

The noteworthy feature of Figure 6 is the weakness of the color gradient
in the disks of these galaxies out to 2 kpc from the plane.
The average over the four galaxies is 0.06 $\pm$ 0.02 mag/kpc
in the sense that the thick disk is redder further from the thin disk at
the 3$\sigma$ level.
Column (2) of Table 2 records the weighted average V--I for the 
region of the disk beyond 900 pcs, and column (3) gives the reddening
in the Milky Way according to NED and Schlegel \etal (1998). The giant branches
of the thick disk regions of these galaxies are consistently red,
with V--I colors ranging from 1.39 to 1.54 (column 4). The bluest color
is that of the least luminous and least massive galaxy, UGCA 442, 
judging by the luminosities and 21 cm linewidths in columns (5) and (6).

In conclusion, there is a general tendency for the giant branch of the stellar
population in the thick disks of these galaxies to show 
color gradients, $\Delta$(V-I)/$\Delta$z, which are zero or slightly positive.
This puts significant constraints on their star formation and chemical enrichment history.

\section{Star formation and chemical enrichment history}

Measurement of metallicity from (V,I) CMDs in stellar populations of the age 
of globular clusters is well understood (Da Costa \& Armandroff 1990, Harris
\& Harris 2002 and references therein). For younger ages, the location of the
RGB depends on both age and metallicity, and it is necessary to refer directly
to the predictions of stellar evolutionary models and synthesize the 
observable quantities from them. To match the observations,
a weighted average V--I color of RGB/AGB stars brighter than M$_I$ =
--3 can be computed from the Padua isochrones (Girardi \etal 2000) for selected
values of the age, t and metallicity, Z of stellar populations. If we assume that
the disk formed stars over an astration time (i.e. time to make stars), which may be short or long
compared with the overall history of the disk, we have the models depicted
in Figure 7. Colors were computed as a sum over four values of Z = (0.0004, 0.001,
0.004, 0.008) with a starting age of 12.6 Gyrs and with convective overshooting assumed. These models show weak dependence of color on timescale and
a much stronger dependence on the final metallicity which the disk attains.

The colors of these models match the colors we observe for the giant branches of
the thick disks in the 4 galaxies of the sample: they bound the range of colors.
We conclude that in the present sample (average luminosity $\sim$~10$^8~L_\odot$) thick disks reached Z = 0.004 to
Z = 0.008 in an astration time that is unconstrained by the present data.
That is, one sixth to one third solar abundance. The mean heavy element
abundances of the stellar populations in these models are $\bar{Z}$ = 0.0018 and 0.0034, i.e. [Fe/H] = --1.0 and --0.78, respectively.  
The very slight color gradient in the thick disk (
0.06 $\pm$ 0.02 mag/kpc) corresponds to a deficit in heavy elements 
of approximately 30\% at z = 1 kpc relative to z = 2 kpc from the midplane of
the disk, or a stellar population younger by $\sim$50\%. The basis of these
approximate estimates is 

$$\frac{d(V-I)}{dz} = 
{\frac{\partial (V-I)}{\partial \log Z}} \frac{d\log Z}{dz} +
{\frac{\partial (V-I)}{\partial \log t}} \frac{d\log t}{dz} $$

\noindent where the partial derivatives are approximately 0.55 and 0.2 respectively
according to the models.

The models can also be used to diagnose the age of the stellar population
from the CMDs. The behavior of an AGB/RGB ratio formed from the ratio 
within the giant branch color window of stars in the interval (21, 23) to
stars in the interval (24, 25) in I magnitude is shown in Figure 8 as a 
function of age for distances of 3--5 Mpc. The ratio is an effective tracer of stellar populations
between 2 $\times$ 10$^8$ and 2 $\times$ 10$^9$ in age, as demonstrated
in Magellanic Cloud star clusters by Frogel, Blanco, \& Mould (1989). 
Column (7) in Table 2 records this AGB/RGB ratio within 100 pixels ($<$ 200 pcs) of
the plane of each galaxy. As shown in Figure 9, NGC 784 and 1560 have substantial intermediate
age populations confined to a thin disk. These are the only sites in the
disks we have sampled where there are intermediate age populations of significance. Sampling statistics illustrated in Figure 9 limit the sensitivity of this technique
to a height of 1 kpc from the plane.

\section{Discussion}
In a recent
review Wyse (2004) considers five possible models for the origin of thick disks.

1. The flattened stellar halo. According to Gilmore and Wyse (1985), the Milky Way's
thick disk has a distinct chemistry from its halo, but further work, involving fields $\sim$10$^\prime$ from the galaxy, would be required to test the equivalent hypothesis for the present sample.

2. The heated thin disk. The classical source of heating is density inhomogeneities
(Wielen and Fuchs 1985) Two alternate sources of heating are star formation and mergers. However, stars of age 2 $\times$ 10$^8$ to 2 $\times$ 10$^9$ years are confined to the thin disk in
the present sample. This model is not supported by the data, although interactions
with galaxies more than 3 Gyrs ago can certainly not be ruled out as heating sources.

3. Dissipative disk formation. Disk settling on timescales longer than the
enrichment timescale gives rise to metallicity gradients (Burkert \etal 1992)
of order a factor of 2 per kpc from z = 0.5 to 1.5 kpc, and steeper elsewhere.
This model is not supported by the data, which show reverse color gradients close to zero.
It is contrived to suppose that reverse color gradients could arise from increasing dustiness further from the plane, although a better test of that hypothesis than stellar photometry might be deep surface photometry, which would be able to detect individual dust clouds. The models of the previous section imply that age gradients are unable to null out metallicity gradients as steep as the dissipative models predict.

4. Accretion from shredded satellites. This model predicts metal poor thick disks,
characteristic of their dwarf components. For example, adding a 10$^7$ L$_\odot$ thick disk ([Fe/H] $\approx$ --1.9) to a 10$^8$ L$_\odot$ galaxy ([Fe/H] $\approx$ --1.4) would yield a color difference
corresponding to a factor of approximately 3 in Z, according to current evidence on the
mass metallicity relation for galaxies, e.g. Tamura, Hirashita, and Takeuchi (2001).
This model in its simplest form is not supported by the data
which show reverse color gradients close to zero.

5. The primordial thick disk. In this model the thick disk is simply the oldest component
of a total galaxy disk that formed over billions of years. According to Brook \etal (2004)
the formation of the early thick disk may have involved a chaotic period of accretion at high redshift. Accretion from shredded satellites is involved in
this process, but, if the accreted satellites have high gas fractions,
astration during and after the merger, and the consequent enrichment, can
eliminate (fully or partly) the difference between the metallicities of the newly arrived
and original material. Models of this nature should be explored as likely
fits to data of the present kind.

\section{Conclusions}

The thick disk stellar populations of a sample of resolved, edge-on, dwarf
galaxies are old and relatively metal rich (one sixth to one tenth solar
abundance in mean metallicity). The intermediate age population is confined to the thin disk.
Color gradients are very slight, and in fact, barely detected. 
Models involving disk heating or halo flattening are not natural fits
to the data, because they differ in enrichment and age from the observed
stellar population. Models involving dissipative collapse tend to predict an
unobserved chemical gradient. Dwarf merger models are problematic,
if complete astration occurs before accretion.

Rather, with the caveat that the present sample is confined to a low mass
subset, the present data seem to favor
a high redshift or primordial origin to the thick disk, possibly in  a
chaotic gas rich merger environment.
The observations are consistent with the ideas of Dalcanton \& Bernstein
(2002), who infer from surface photometry of a large sample that thick disks
are a standard phase in the early formation of disk galaxies.

Resolution into giant branch CMDs helps separate age and metallicity
effects in analyzing stellar populations. It is therefore desirable to
extend the present study to a larger sample of less dwarfish galaxies
using ACS on HST.

\acknowledgments
I would like to thank the School of Physics of the University of Melbourne
for their hospitality while the first draft was being written. 
I also acknowledge the help of an anonymous referee whose comments
improved the paper materially. This research has
made use of NED, which is operated by JPL for NASA, and the HST archive which is operated by STScI for NASA. Images were processed with IRAF which is maintained by NOAO. NOAO and STScI are managed by AURA for the agencies.

\vskip 1 truein
\appendix
\centerline{\bf Appendix}
Although accurate distances are not essential to this investigation, 
it is of interest to examine the effects of incompleteness on the TRGB
distances reported in Table 1. For this purpose in both filters, we added 
200 stars per chip to each of the 4 galaxies for which CMDs have been
obtained, employing a uniform spatial and magnitude distribution in the range
22.8 $<$ I $<$ 25.8 mag (0.75 mag fainter in NGC 1560 and UGCA 442). 
The measured incompleteness is displayed in Figure 10.

Because the RGB tip lies in a region where completeness declines slowly
and linearly with increasing magnitude, completeness corrections are 
negligible relative to the uncertainties arising from limited knowledge
of reddening and metallicity. The color gradients measured in $\S$3
were also found to be robust against incompleteness.

\pagebreak
\centerline{\bf Figure Captions}





\begin{figure}[h]
\figurenum{2}
\centering \leavevmode
\vskip .5 truein
\hskip -.7 truein
\epsfysize 7 truein
\epsfbox{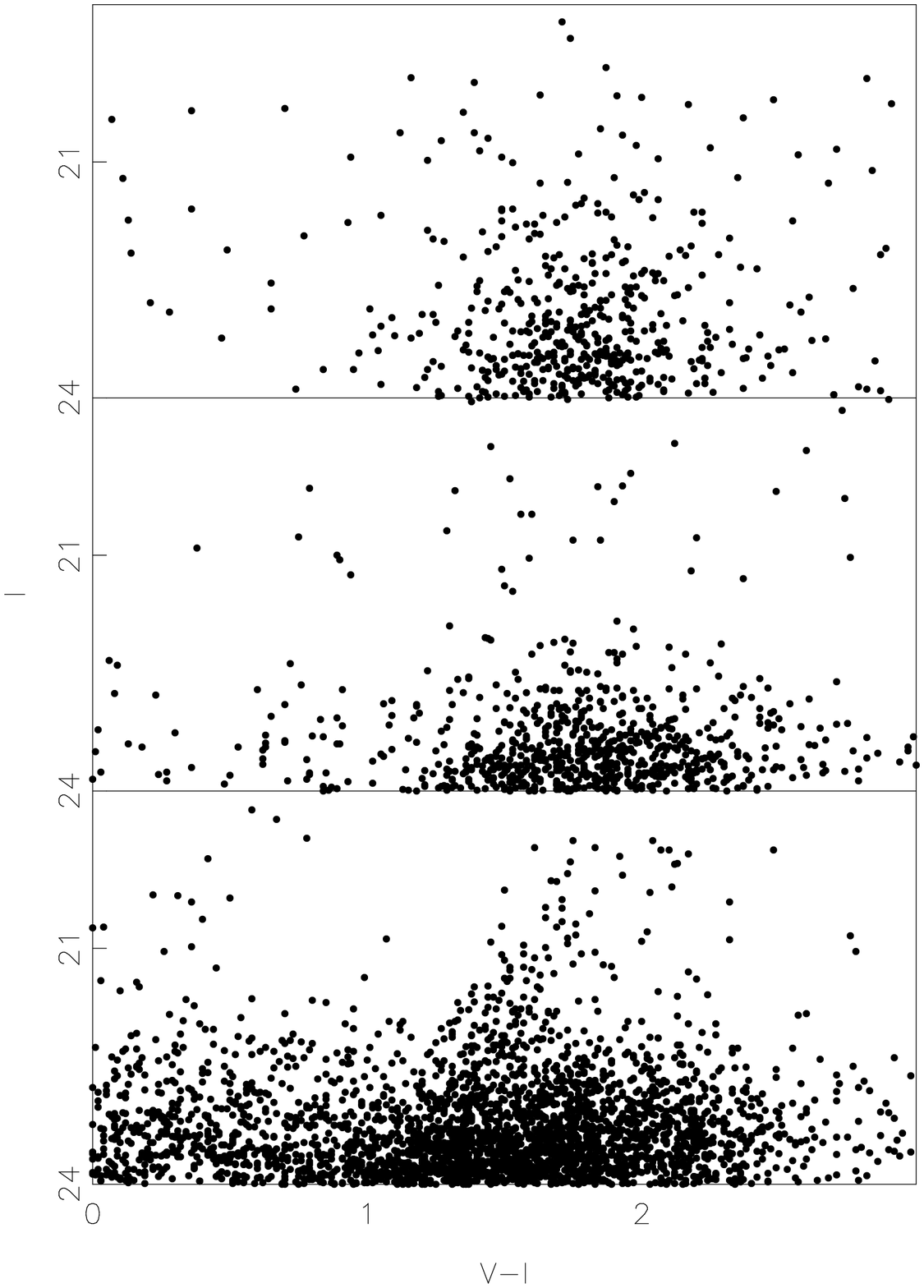}
\figcaption{ {\it (a) CMD for NGC 784. 
Bottom panel -- stars within 200 pixels of the midplane of the galaxy.
Top panel -- stars more than 500 pixels from the midplane of the galaxy.
The middle panel is those in between.}} 
\end{figure}
\clearpage

\begin{figure}[h]
\figurenum{2}
\centering \leavevmode
\hskip -.7 truein
\epsfysize 7 truein
\epsfbox{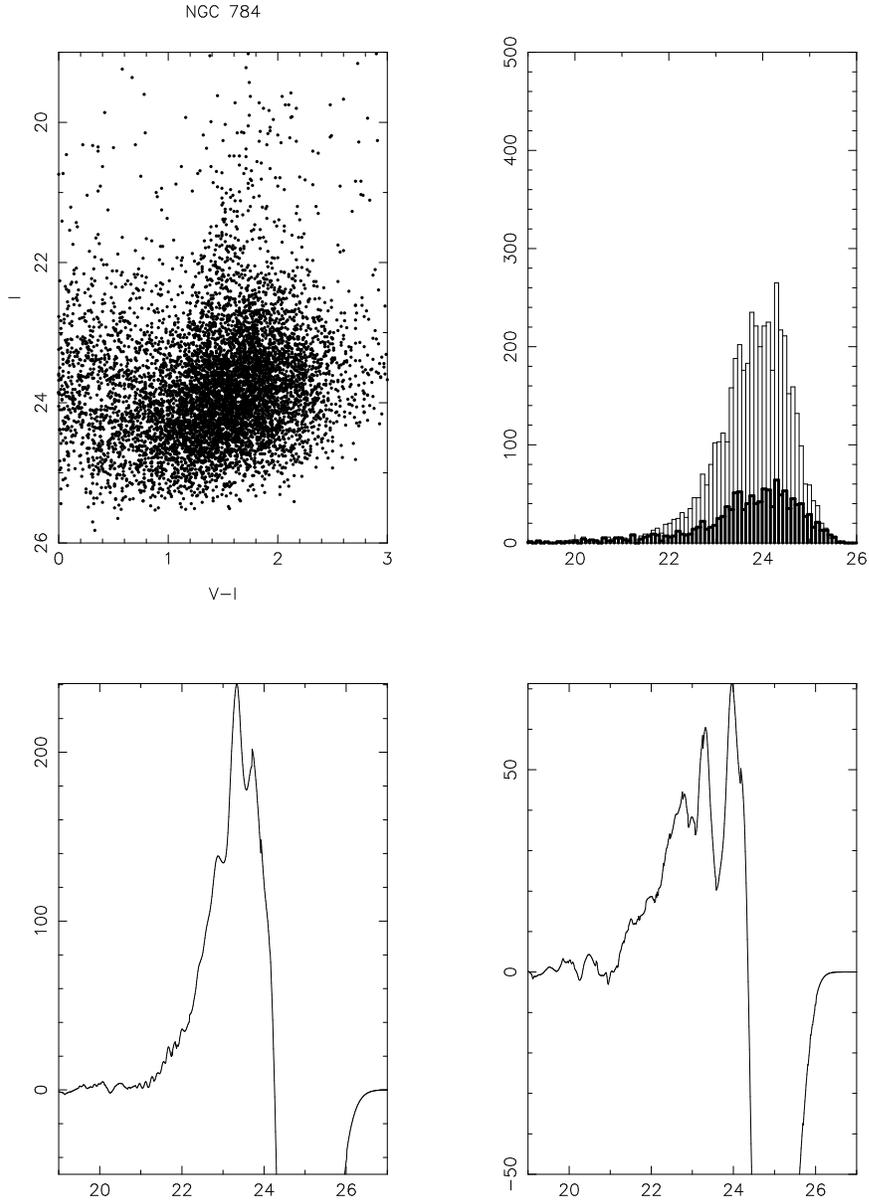}
\figcaption{ {\it (b) CMD for NGC 784. Top left -- all stars.
Top right -- the I band giant branch luminosity function for 1 $<~V-I~<$ 2; 
the lower bold histogram
refers to those stars more than 200 pixels from the midplane of the galaxy.
Bottom left -- the Sakai \etal (1996) edge detector for the tip of the RGB.
Bottom right -- those stars more than 200 pixels from the midplane of the galaxy.}} 
\end{figure}
\clearpage

\begin{figure}[h]
\figurenum{3}
\centering \leavevmode
\vskip .5 truein
\hskip -.7 truein
\epsfysize 7 truein
\epsfbox{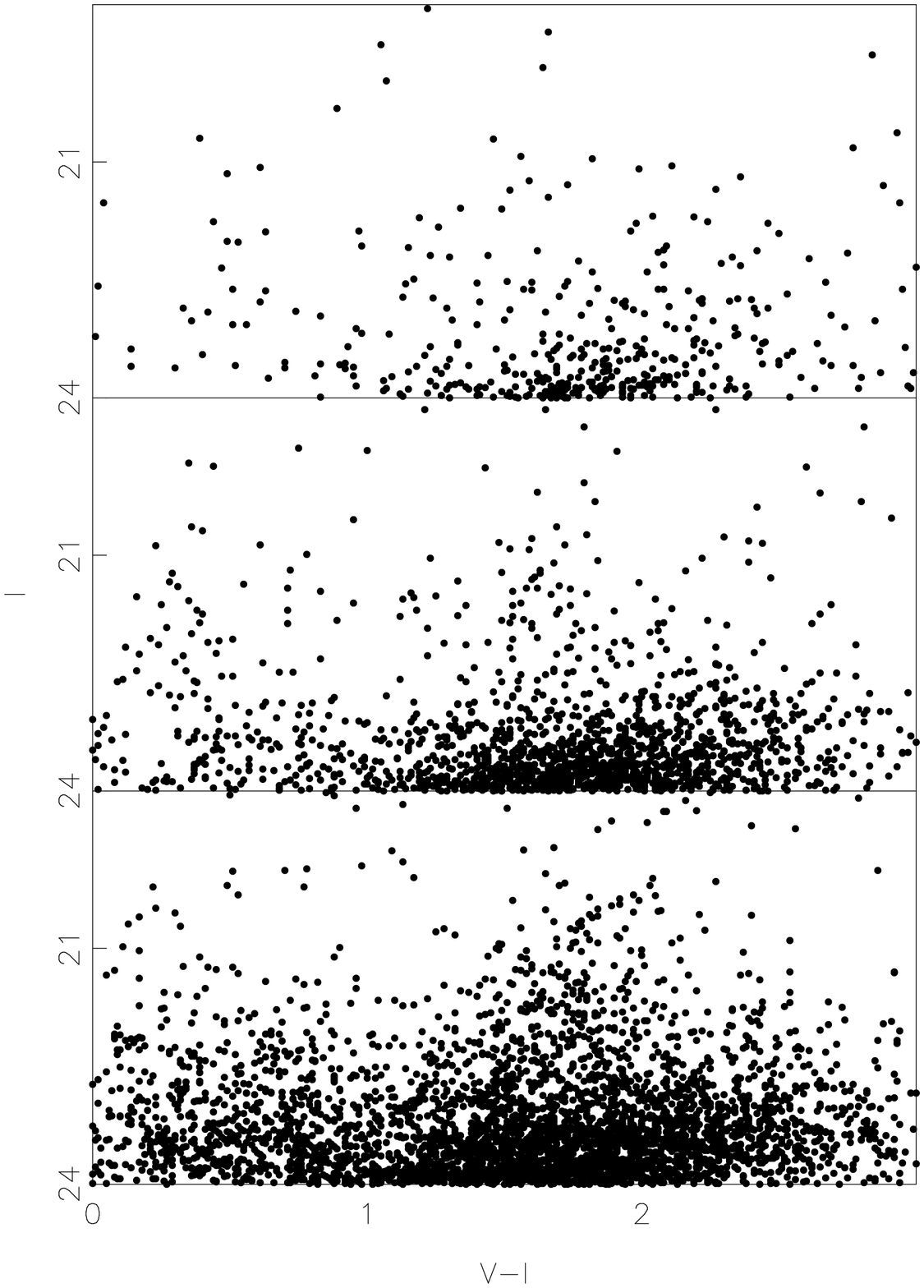}
\figcaption{ {\it (a) CMD for NGC 1560. The figure key is the same as Figure 2a.}}
\end{figure}
\clearpage

\begin{figure}[h]
\figurenum{3}
\centering \leavevmode
\vskip .5 truein
\hskip -.7 truein
\epsfysize 7 truein
\epsfbox{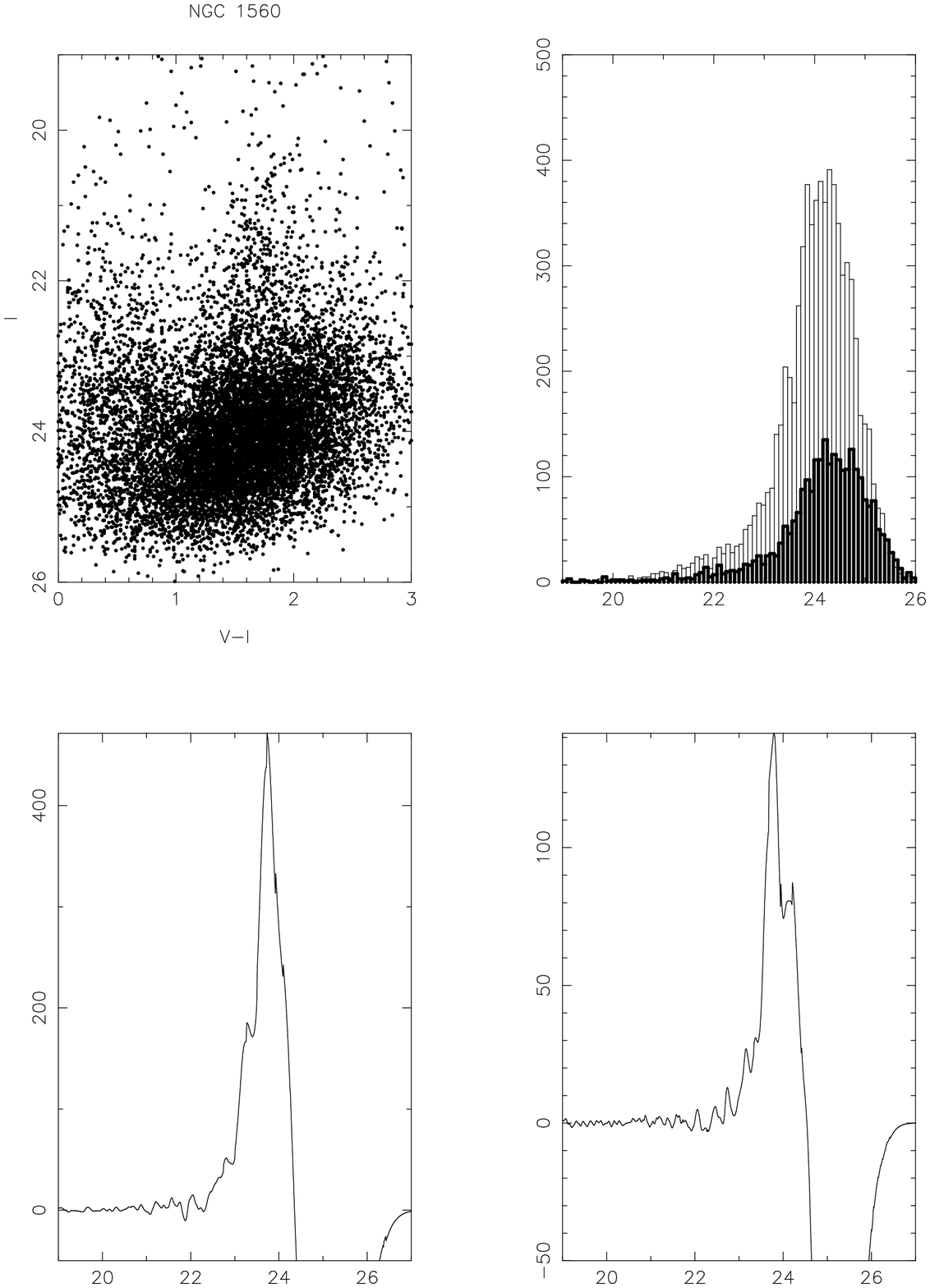}
\figcaption{ {\it (b) CMD for NGC 1560. The figure key is the same as Figure 2b.}}
\end{figure}
\clearpage

\begin{figure}[h]
\figurenum{4}
\centering \leavevmode
\vskip .5 truein
\hskip -.7 truein
\epsfysize 7 truein
\epsfbox{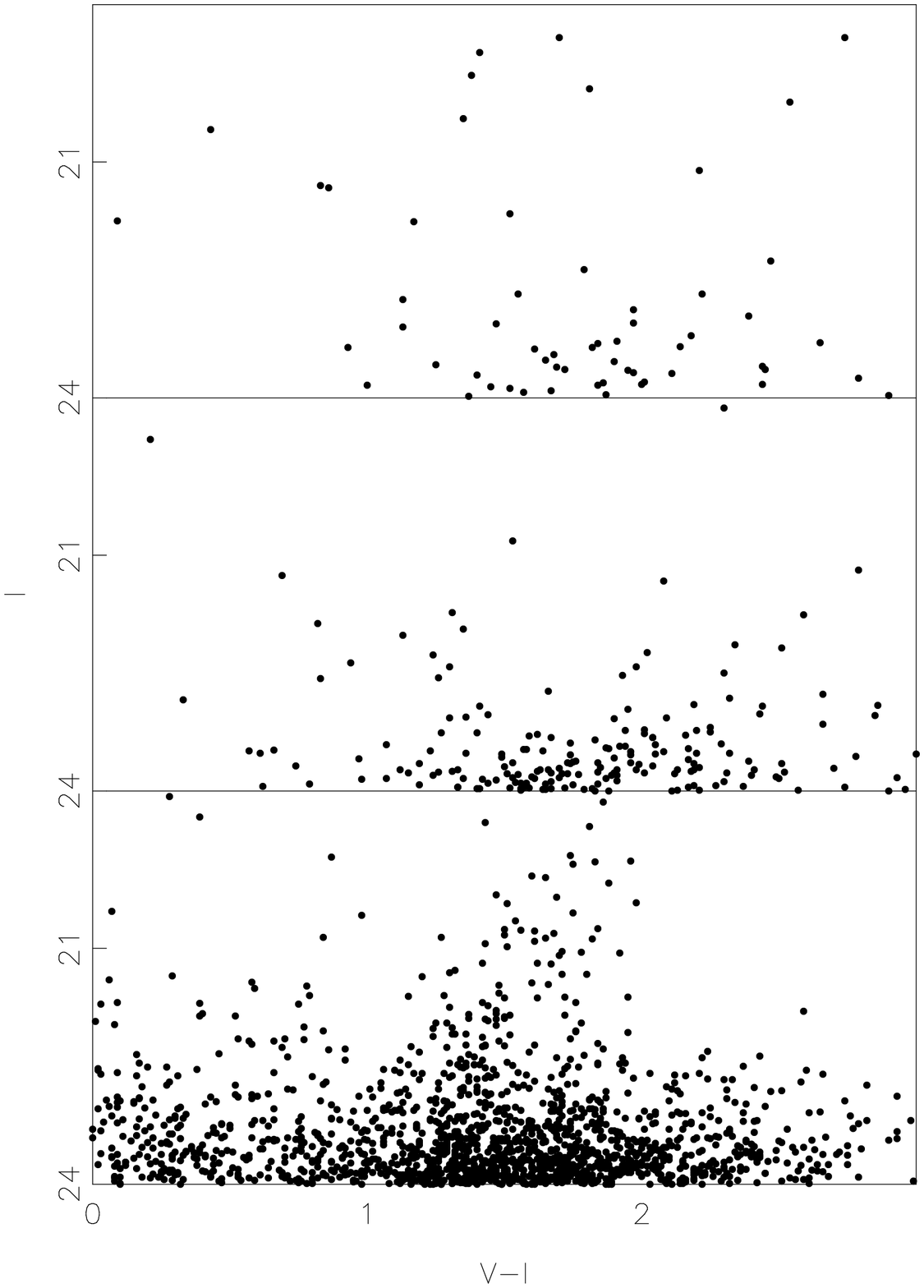}
\figcaption{ {\it (a) CMD for UGC 1281. The figure key is the same as Figure 2a.}}
\end{figure}
\clearpage

\begin{figure}[h]
\figurenum{4}
\centering \leavevmode
\vskip .5 truein
\hskip -.7 truein
\epsfysize 7 truein
\epsfbox{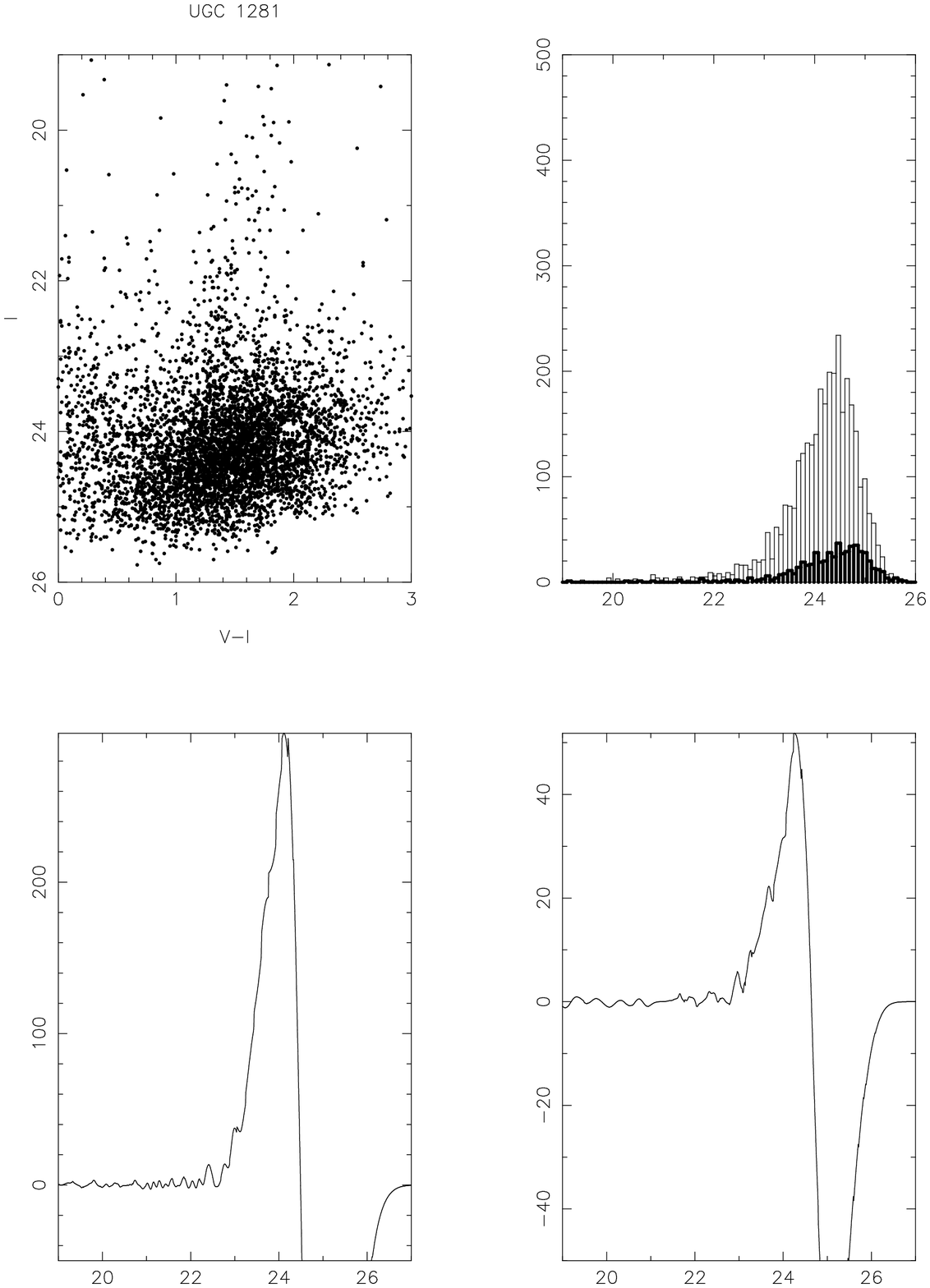}
\figcaption{ {\it (b) CMD for UGC 1281. The figure key is the same as Figure 2b.}}
\end{figure}
\clearpage

\begin{figure}[h]
\figurenum{5}
\centering \leavevmode
\vskip .5 truein
\hskip -.7 truein
\epsfysize 7 truein
\epsfbox{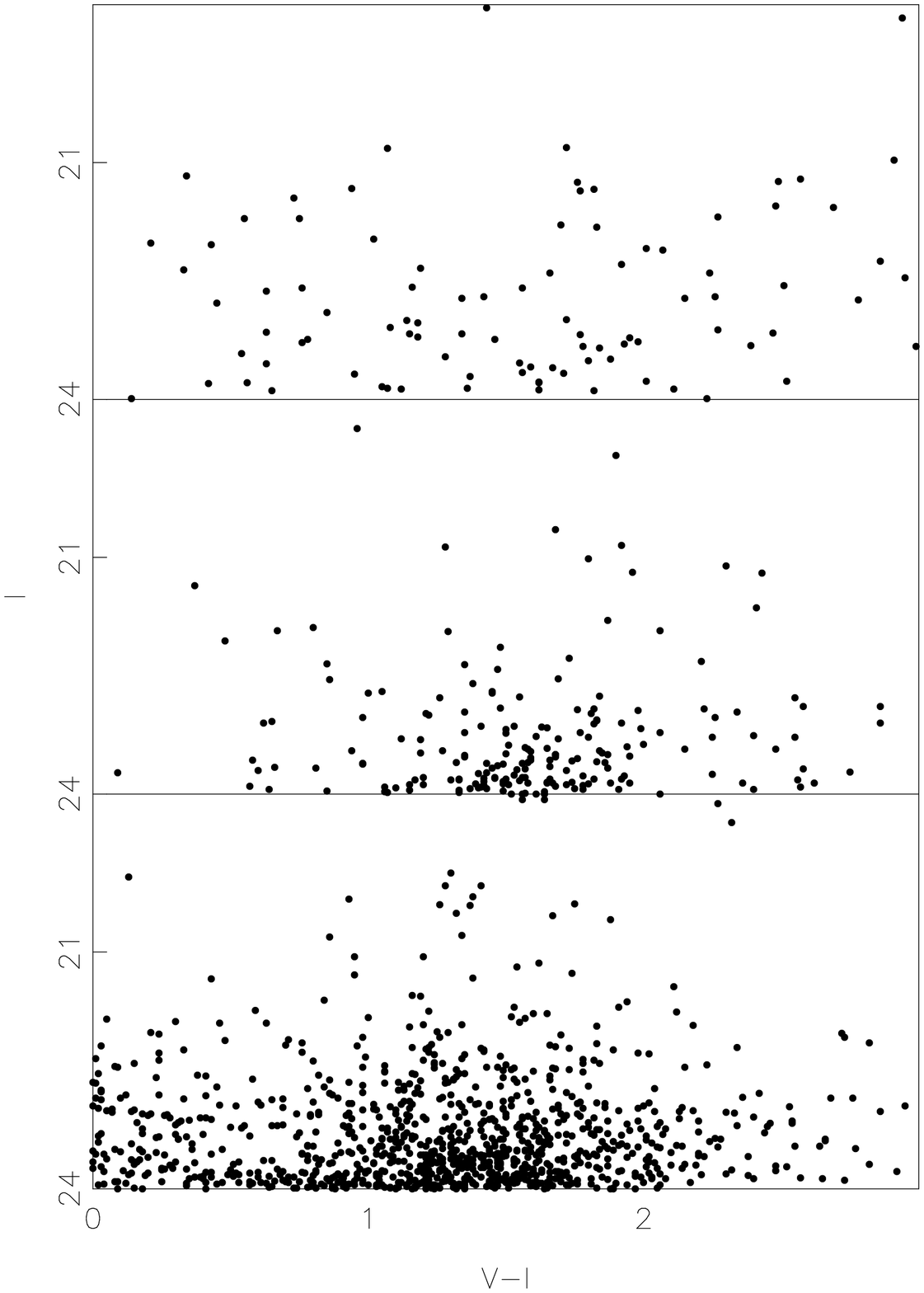}
\figcaption{ {\it (a) CMD for UGCA 442. The figure key is the same as Figure 2a.}}
\end{figure}
\clearpage

\begin{figure}[h]
\figurenum{5}
\centering \leavevmode
\vskip .5 truein
\hskip -.7 truein
\epsfysize 7 truein
\epsfbox{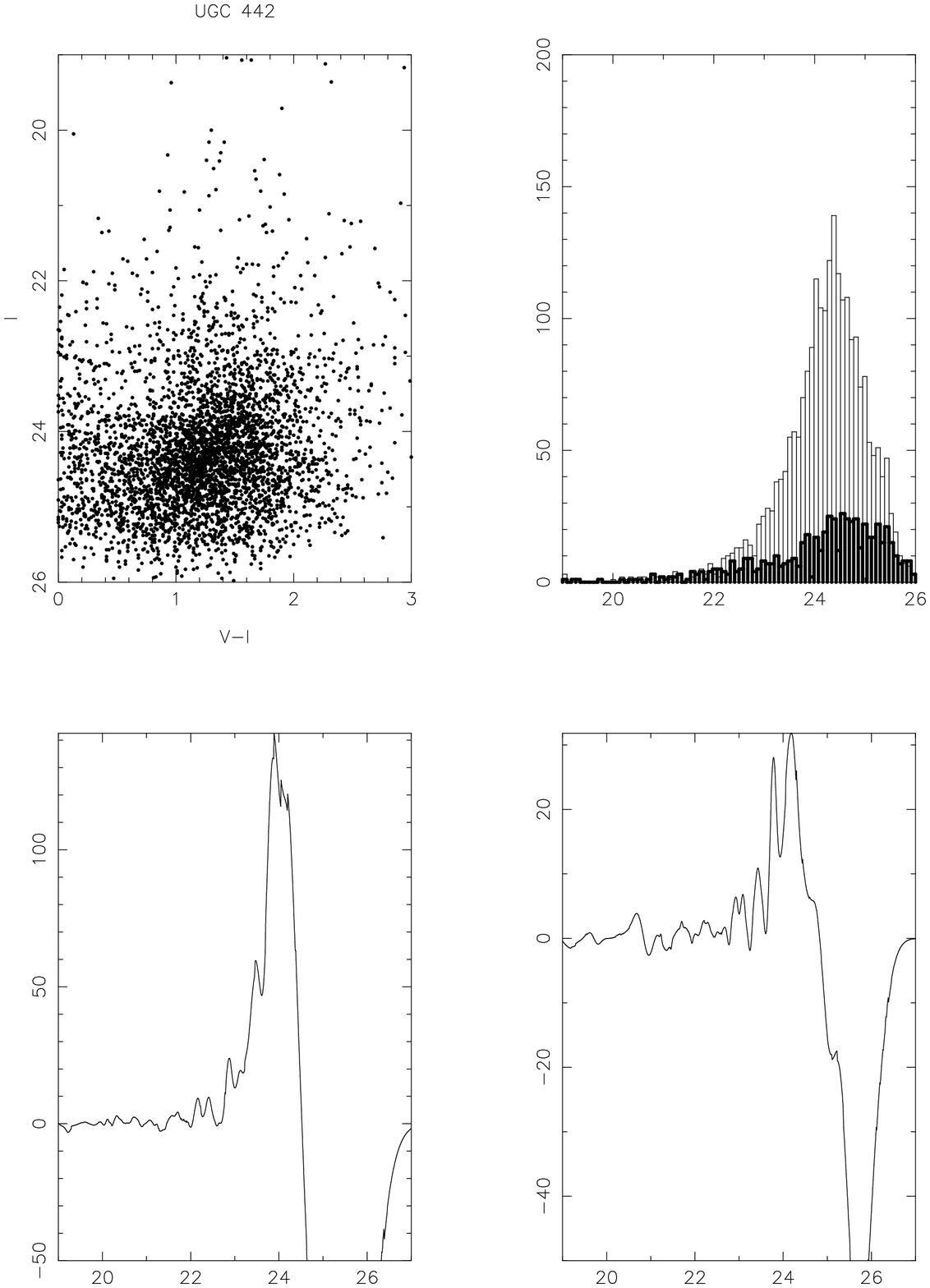}
\figcaption{ {\it (b) CMD for UGCA 442. The figure key is the same as Figure 2b.}}
\end{figure}
\clearpage

\begin{figure}[h]
\figurenum{6}
\centering \leavevmode
\vskip .5 truein
\hskip -.7 truein
\epsfysize 5 truein
\epsfbox{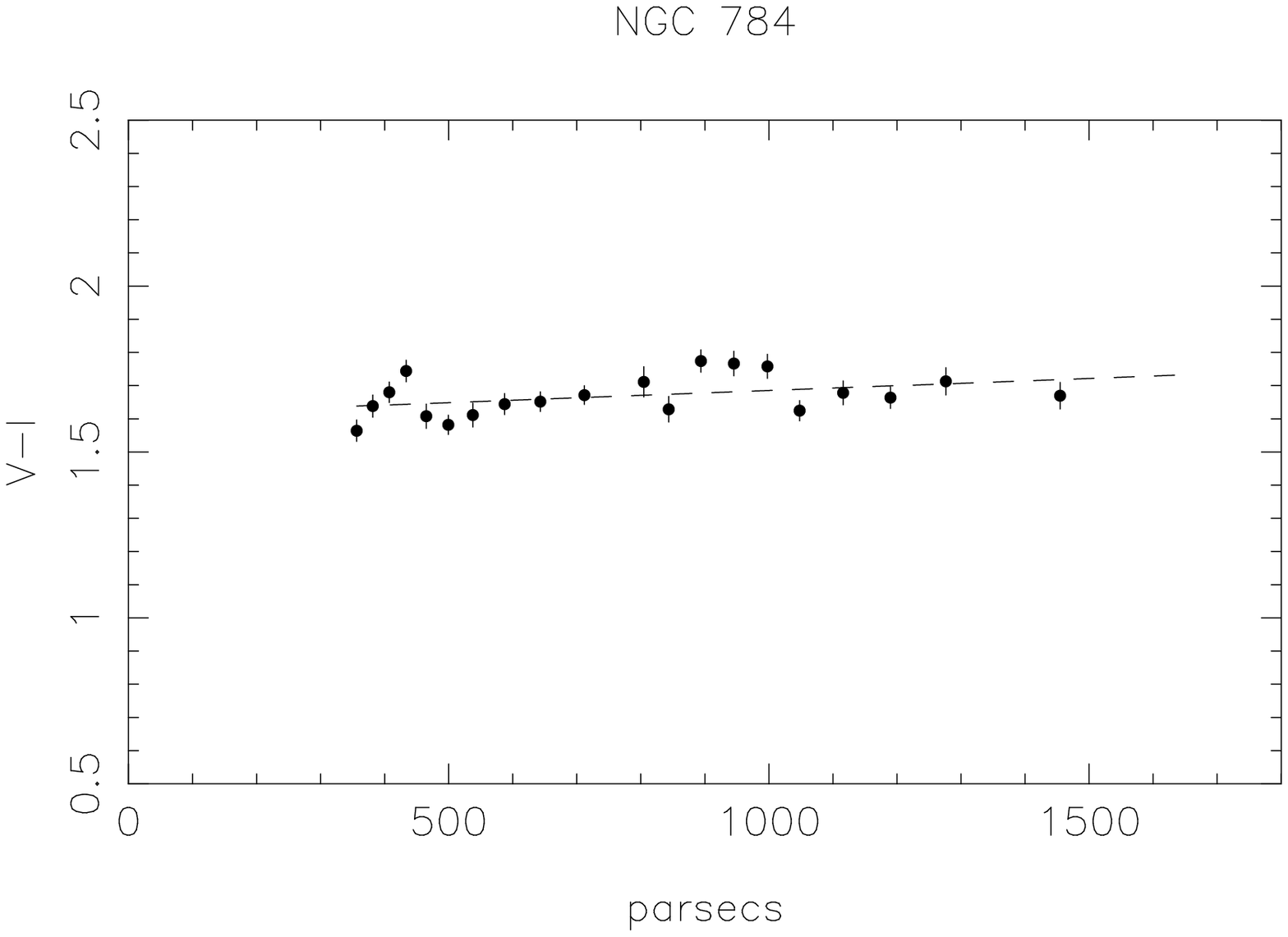}
\figcaption{ {\it The mean color of the giant branch of each galaxy as a function
of projected distance from the plane of the disk.
(a) NGC 784.}}
\end{figure}
\clearpage

\begin{figure}[h]
\figurenum{6}
\centering \leavevmode
\vskip .5 truein
\hskip -.7 truein
\epsfysize 5 truein
\epsfbox{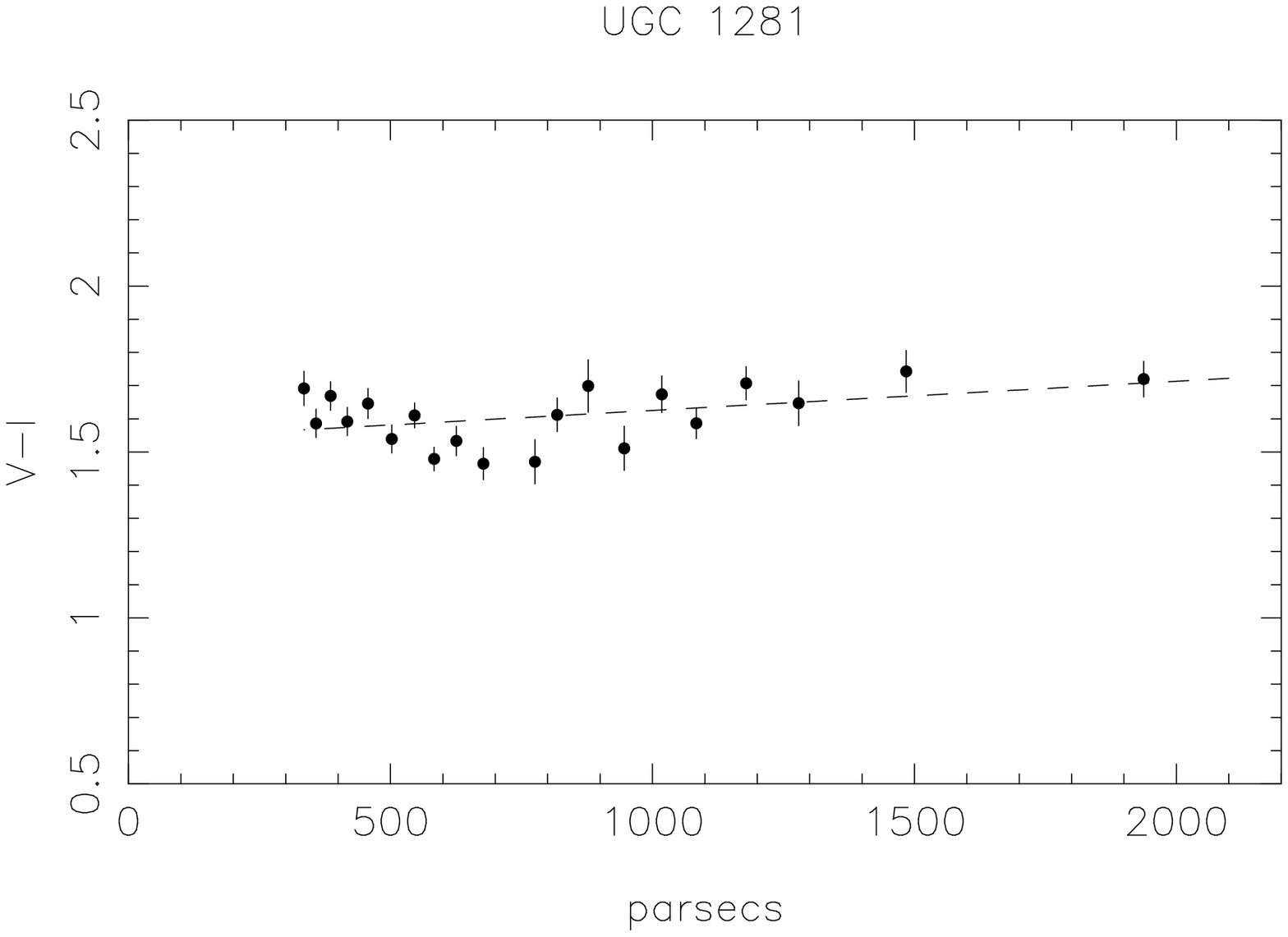}
\figcaption{ {\it (b) UGC 1281.}}
\end{figure}
\clearpage

\begin{figure}[h]
\figurenum{6}
\centering \leavevmode
\vskip -.75 truein
\hskip -.7 truein
\epsfysize 5 truein
\epsfbox{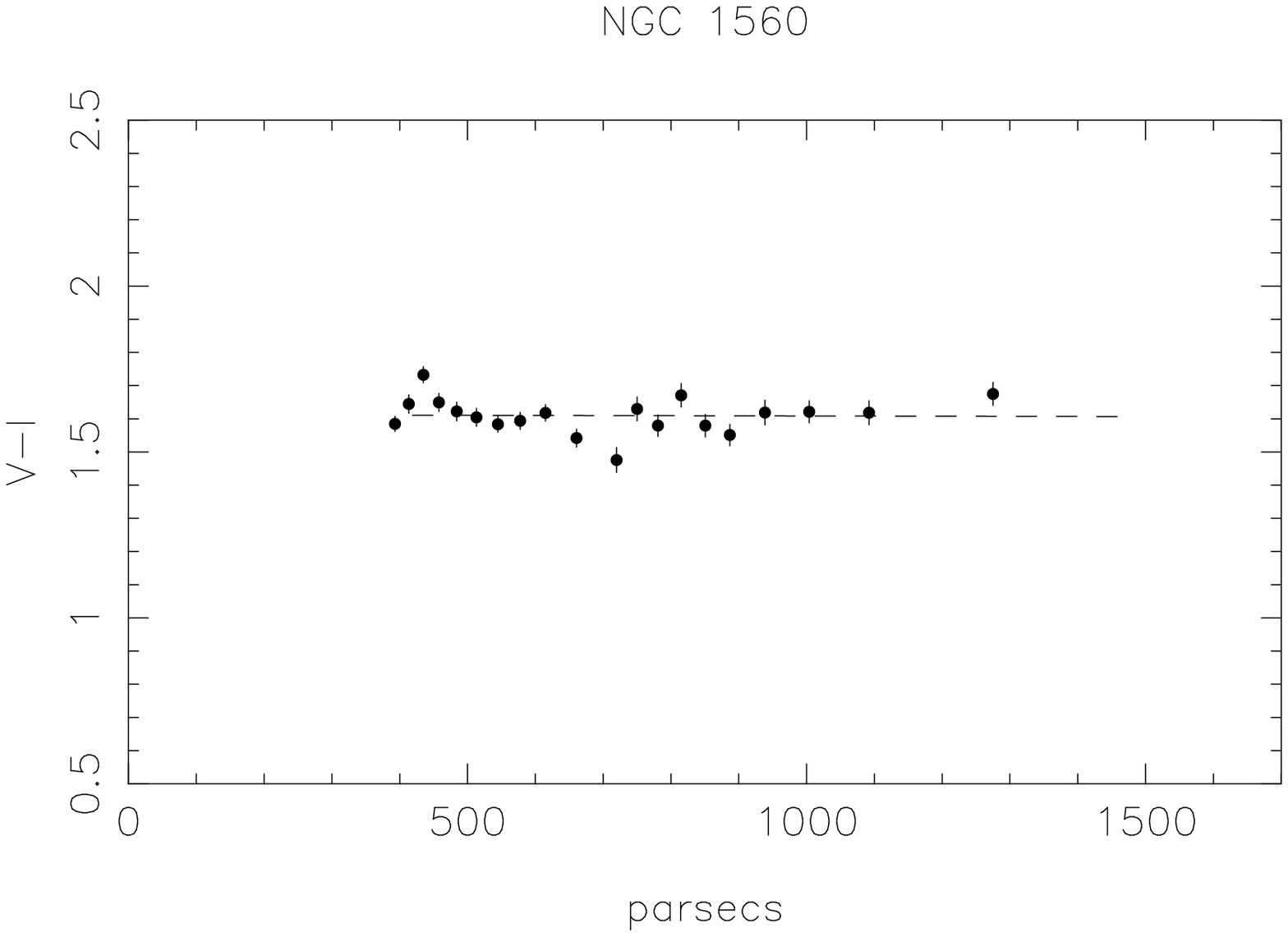}
\figcaption{ {\it (c) NGC 1560.}}
\end{figure}
\clearpage

\begin{figure}[h]
\figurenum{6}
\centering \leavevmode
\vskip -.25 truein
\hskip -.7 truein
\epsfysize 5 truein
\epsfbox{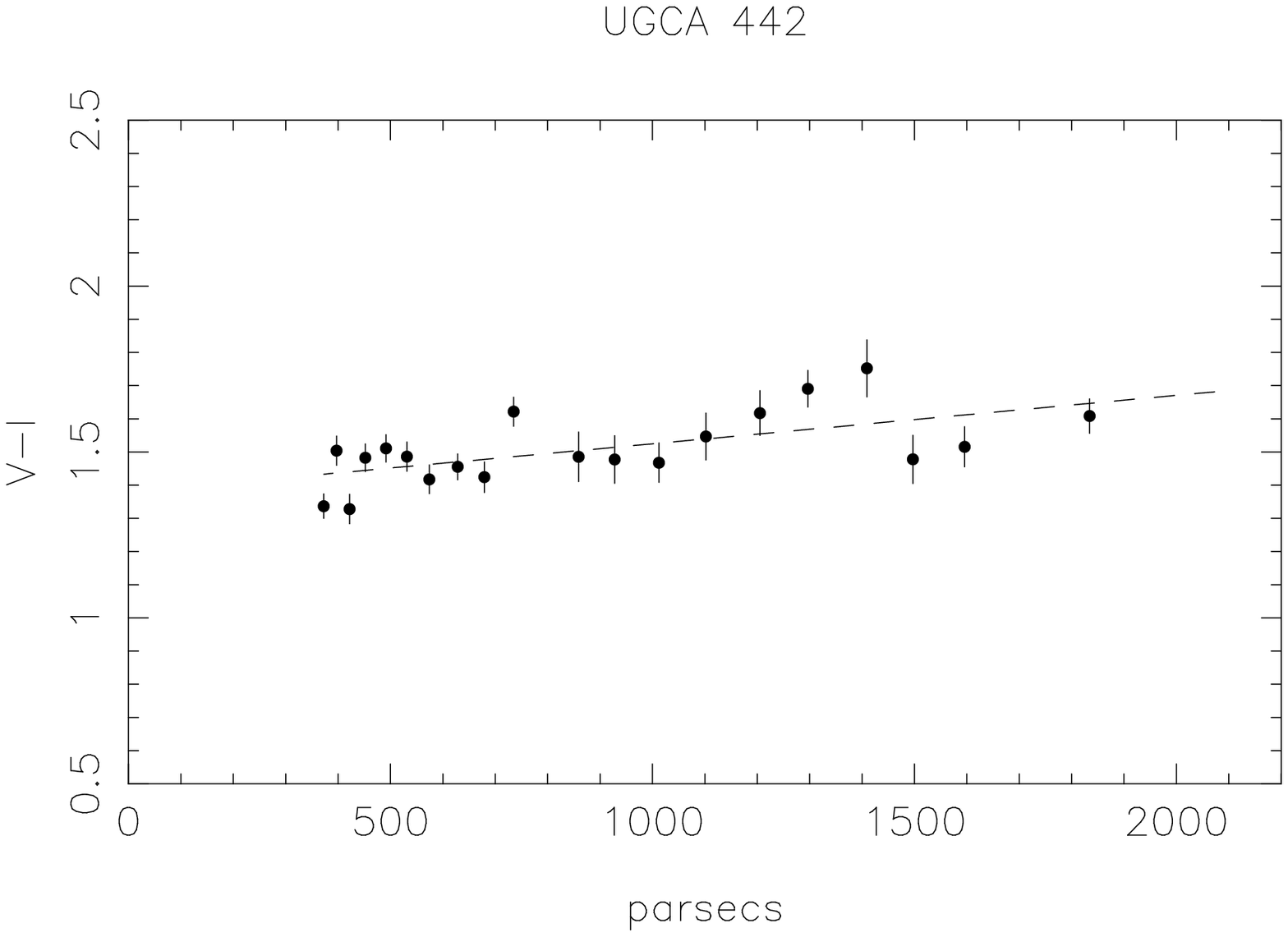}
\figcaption{ {\it (d) UGCA 442.}}
\end{figure}
\clearpage

\begin{figure}[h]
\figurenum{7}
\centering \leavevmode
\vskip -.25 truein
\hskip -.7 truein
\epsfysize 4 truein
\epsfbox{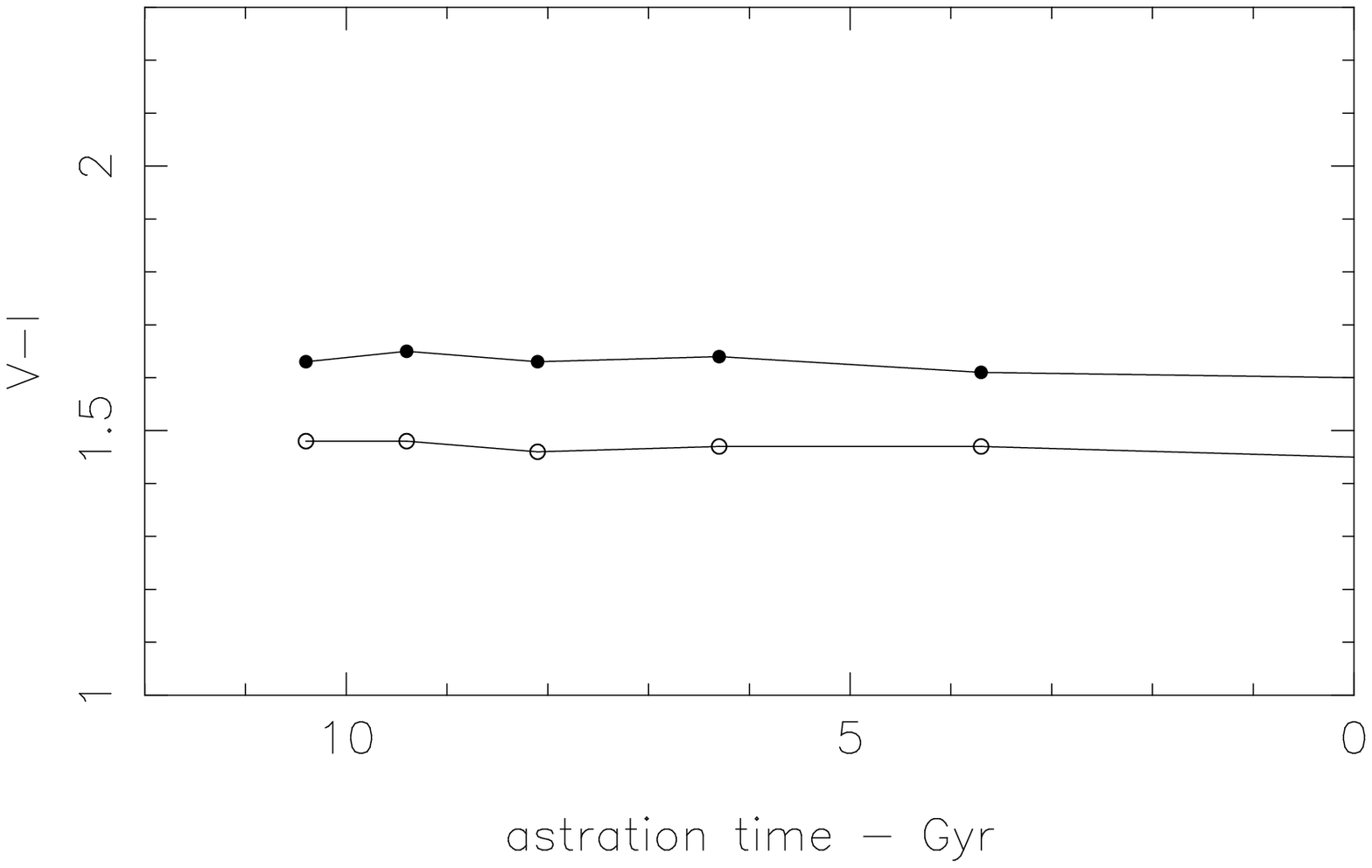}
\figcaption{ {\it Mean V--I color of the mean giant branch with M$_I~<$ --3 mag
in the Padua isochrones as a function of the time to enrich
the stellar population from Z = 0.0004 to Z = 0.008 (solid symbols)
or to Z = 0.004 (open symbols).}}
\end{figure}
\clearpage

\begin{figure}[h]
\figurenum{8}
\centering \leavevmode
\vskip -.25 truein
\hskip -.7 truein
\epsfysize 5 truein
\epsfbox{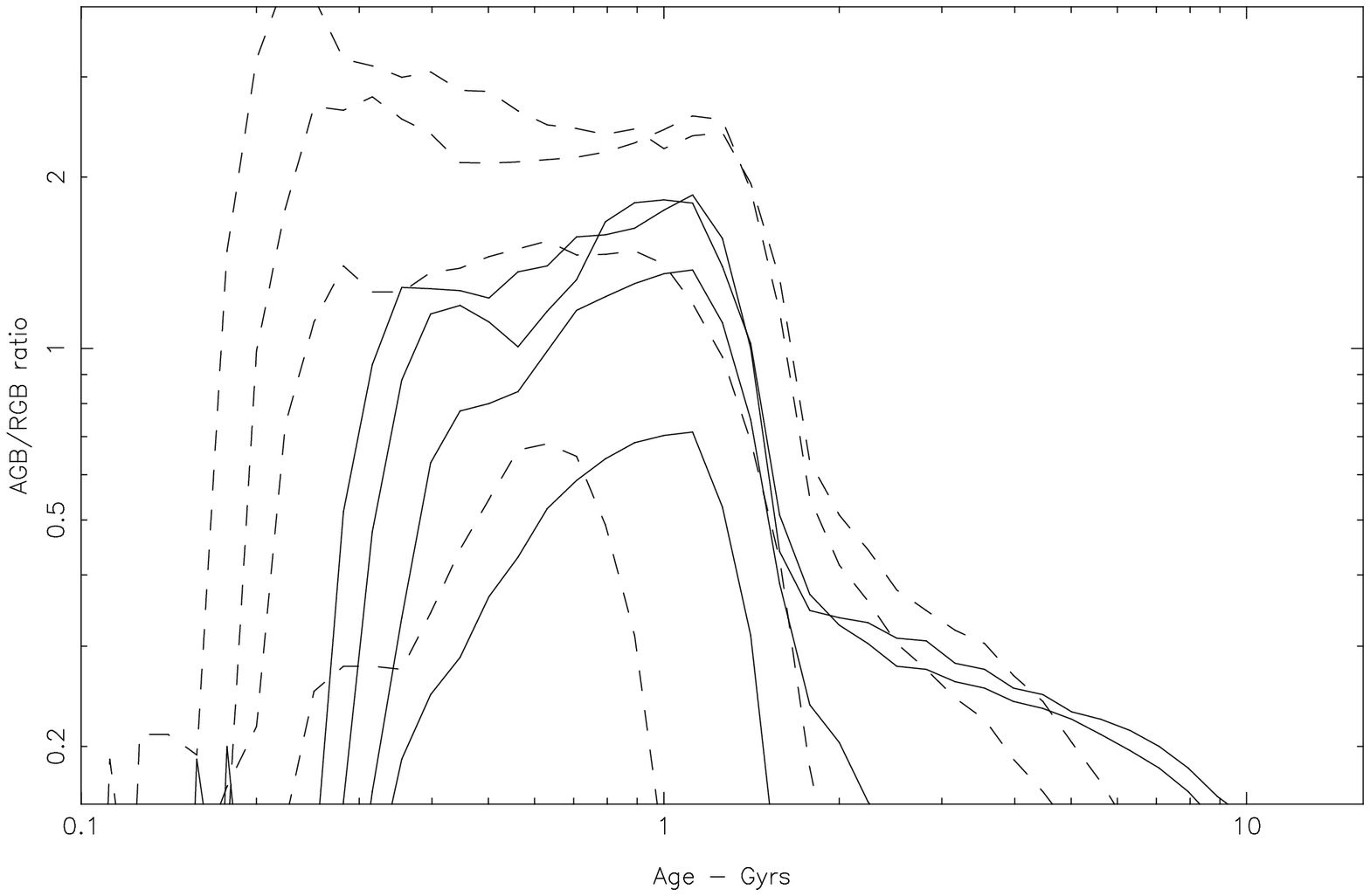}
\figcaption{ {\it The ratio of upper AGB stars (21 $<$ I $<$ 23) to RGB
tip stars (24 $<$ I $<$ 25) in Padua model populations as a function of age.
Solid line (m-M = 27.5 mag). Dashed line (m-M = 28.2 mag). The four curves
at each modulus are for metallicities Z = 0.0004, 0.001, 0.004, and 0.008.}}
\end{figure}
\clearpage

\begin{figure}[h]
\figurenum{9}
\centering \leavevmode
\vskip -.25 truein
\hskip -.7 truein
\epsfysize 5 truein
\epsfbox{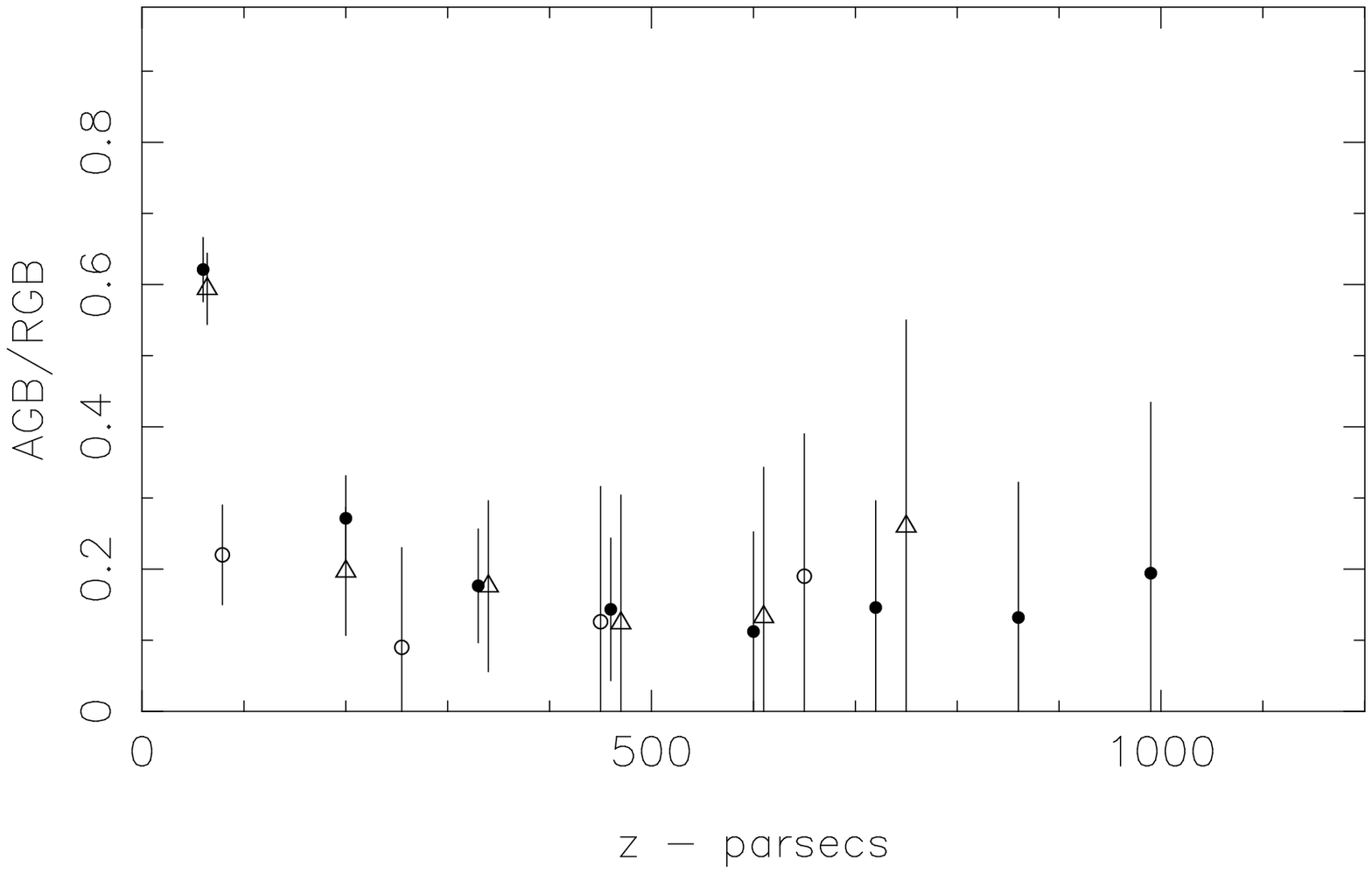}
\figcaption{ {\it AGB/RGB ratio as a function of distance from the plane.
Solid symbol: NGC 1560, open circle: UGCA 442, triangle: NGC 784.
The intermediate age population is confined to the plane in NGC 784 and 1560.
UGC 1281 is not shown, but is similar to UGCA 442.}}
\end{figure}
\clearpage

\begin{figure}[h]
\figurenum{10}
\centering \leavevmode
\vskip -.25 truein
\hskip -.7 truein
\epsfysize 7 truein
\epsfbox{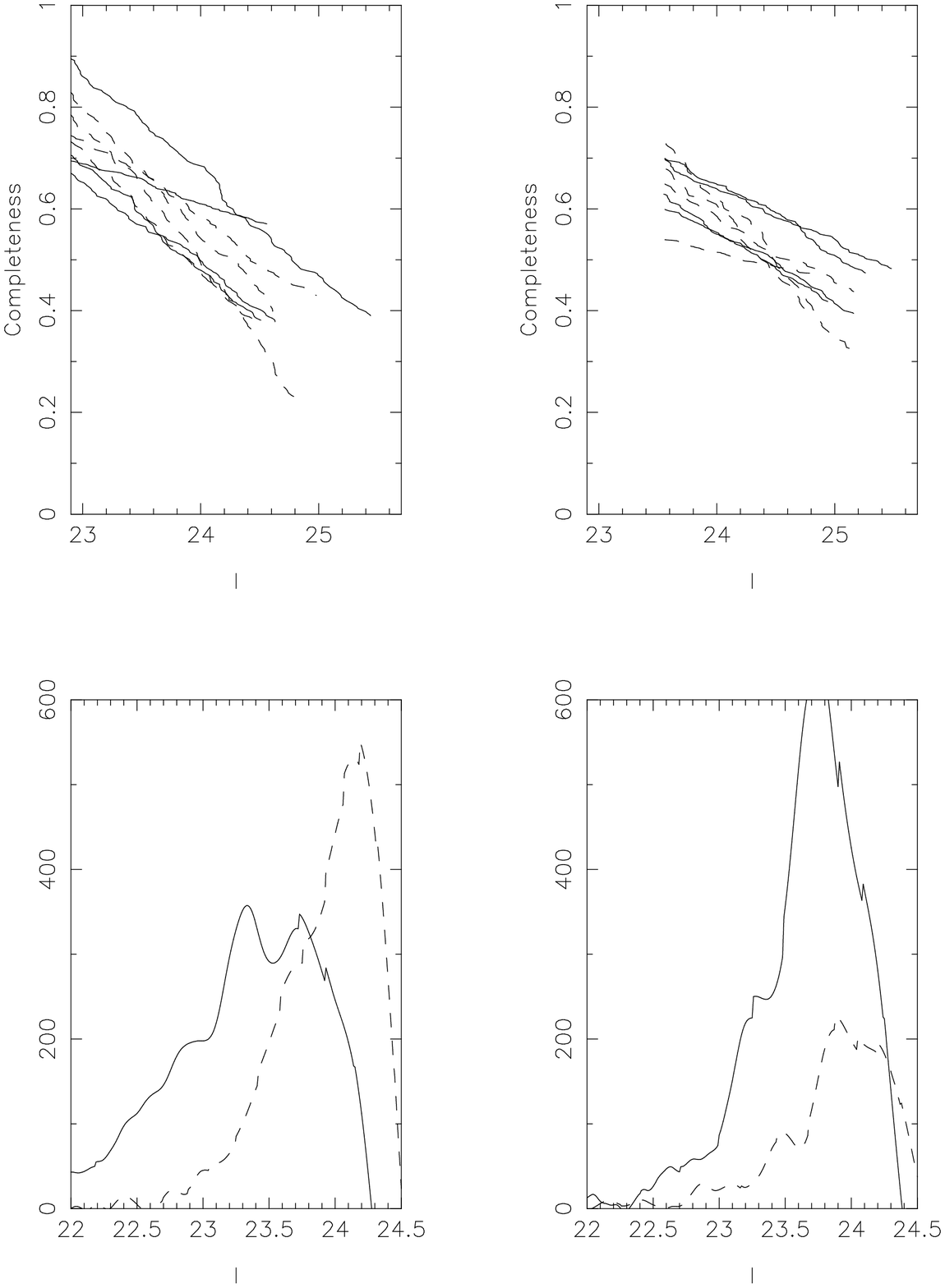}
\figcaption{ {\it Completeness corrected edge detector for the tip of the RGB 
(lower panels). Fake star experiments allow the completeness to be estimated,
and this is shown chip-by-chip in the upper panels. The left hand panels
are NGC 784 (solid) and UGC 1281 (dashed). The right hand panels
are NGC 1560 (solid) and UGCA 442 (dashed). The completeness corrected TRGB
edge detector is displayed with the same galaxy key. }}
\end{figure}

\clearpage

\centerline{\bf Table 1: WFPC2 observations}

\begin{tabbing}
NGC 1560sssss \= b/a ssssssss \= Filters ssssssssssss \= Exposure time \= absorption \= \kill
Galaxy   \> Axial ratio \> Filters       \> Exposure time \> A$_I$\>(m-M)$_I$\\
 (1)      \> (2)   \>      (3)           \> (4)           \> (5)\> (6)\\
 NGC 1560 \> 0.16 \> F606W, F814W  \> 600s            \>0.365\> 27.6$\pm$0.15\\
 UGC 1281 \> 0.11 \> F606W, F814W  \> 2 $\times$ 300s \>0.09 \> 28.1$^{+0.35}_{-0.15}$\\
 NGC 784  \> 0.23 \> F606W, F814W  \> 2 $\times$ 300s \>0.115\> 27.4$^{+0.75}_{-0.15}$\\
 UGCA 442 \> 0.14 \> F606W, F814W  \> 600s            \>0.032\> 27.9$^{+0.35}_{-0.15}$\\
\end{tabbing}

\leftline{Column notes:}
\noindent (2) and (5) Source: NED.\\
(5) and (6) Magnitudes.\\
(6) Apparent distance moduli derived in section 2.

\pagebreak

\centerline{\bf Table 2: Giant branch properties}

\begin{tabbing}
NGC 1560ssss \= b/asss \= Filterss \= Exposure  \= absorption \= sssssssss \= sssssssssssss \= \kill
Galaxy   \> V--I \> E(V--I) \> $(V-I)_0$\> M$_V$\>w$_{20}$\> AGB/RGB\>d(V-I)/dz\\
 (1)      \> (2)   \>      (3)     \> (4) \> (5)\> (6)\> (7) \> (8)\\
 NGC 1560 \> 1.71 \> 0.26  \>  1.45\>--16.2\> 157\> 0.56\> 0.01$\pm$0.05\\
 UGC 1281 \> 1.60 \> 0.06  \>  1.54\>      \> 129\> 0.15\> 0.09$\pm$0.04\\
 NGC 784  \> 1.64 \> 0.08  \>  1.56\>--15.8\> 116\> 0.67\> 0.06$\pm$0.04\\
 UGCA 442 \> 1.41 \> 0.02  \>  1.39\>--14.8\> 105\> 0.21\> 0.11$\pm$0.10\\
\end{tabbing}

\leftline{Column notes:}
\noindent (2) Average giant branch color derived in section 3.\\ 
(3) Reddening. Source: NED.\\
(4) Average giant branch color corrected for reddening.\\
(5) Absolute visual magnitude of the galaxy. Source of apparent magnitude: NED.\\
(6) Velocity width in km s$^{-1}$ at 20\% of the 21cm peak. Source: NED.\\
(7) Ratio of stars above/below the TRGB (section 4). Uncertainties 0.01--0.02.\\
(8) Color gradient in Figure 6 in magnitudes/kpc.

\end{document}